\documentclass[12pt,preprint]{aastex}
\usepackage{xcolor}
\usepackage{soul}

\begin{document}

\title{Peculiar Transverse Velocities of Galaxies from Quasar Microlensing. Tentative Estimate of the Peculiar Velocity Dispersion at $z\sim 0.5$  }

\author{E. MEDIAVILLA\altaffilmark{1,2}, J. JIM\'ENEZ-VICENTE\altaffilmark{3,4}, J. A. MU\~NOZ\altaffilmark{5,6}, E. BATTANER\altaffilmark{3,4}}

\altaffiltext{1}{Instituto de Astrof\'{\i}sica de Canarias, V\'{\i}a L\'actea S/N, La Laguna 38200, Tenerife, Spain}
\altaffiltext{2}{Departamento de Astrof\'{\i}sica, Universidad de la Laguna, La Laguna 38200, Tenerife, Spain}
\altaffiltext{3}{Departamento de F\'{\i}sica Te\'orica y del Cosmos, Universidad de Granada, Campus de Fuentenueva, 18071 Granada, Spain}
\altaffiltext{4}{Instituto Carlos I de F\'{\i}sica Te\'orica y Computacional, Universidad de Granada, 18071 Granada, Spain}
\altaffiltext{5}{Departamento de Astronom\'{\i}a y Astrof\'{\i}sica, Universidad de Valencia, 46100 Burjassot, Valencia, Spain.}
\altaffiltext{6}{Observatorio Astron\'omico, Universidad de Valencia, E-46980 Paterna, Valencia, Spain}

\begin{abstract}
We propose to use the flux variability of lensed quasar images induced by gravitational microlensing  to measure the transverse peculiar velocity of lens galaxies over a wide range of redshift. Microlensing variability is caused by the {  motions of the observer, the lens galaxy  (including the motion of the stars within the galaxy), and the source};
hence, its frequency is directly related to the galaxy's transverse peculiar velocity. The idea is to count time-event rates (e.g., peak or caustic crossing rates) in the observed microlensing light curves of lensed quasars that can be compared with model predictions for different values of the transverse peculiar velocity. To compensate for the large time-scale of microlensing variability
we propose to count and model the number of events in an ensemble of gravitational lenses. We develop the methodology to achieve this goal and apply it to an ensemble of 17 lensed quasar systems
. In spite of the shortcomings of the available data,
we have obtained tentative estimates of the peculiar velocity dispersion of lens galaxies at $z\sim 0.5$, $\sigma_{\rm pec}(0.53\pm0.18)\simeq(638\pm213)\sqrt{\langle m \rangle/0.3 M_\odot} \, \rm km\, s^{-1}$. { Scaling at zero redshift we derive, $\sigma_{\rm pec}(0)\simeq(491\pm164) \sqrt{\langle m \rangle/0.3 M_\odot} \, \rm km\, s^{-1}$, consistent with peculiar motions of nearby galaxies} and with recent $N$-body nonlinear reconstructions of the Local Universe based on $\Lambda$CDM. We analyze the different sources of uncertainty of the method and find that for the present ensemble of 17 lensed systems the error is dominated by Poissonian noise, but that for larger ensembles the impact of the uncertainty on the average stellar mass may be significant.

\end{abstract}


\section{Introduction}

The motion of galaxies with respect to the smooth Hubble flow, i.e., the peculiar velocity field of galaxies, is a useful probe of cosmology and galaxy formation. Peculiar velocities allow us to trace the overall matter distribution (including dark matter) over a wide range of scales. On cosmological scales the coherent flows of galaxies toward overdense regions are determined by the overall amount of matter and its large scale distribution (Kaiser 1988). Consequently,  galaxy peculiar velocities provide powerful tests of the cosmological model through measurements of the linear growth rate; these measurements are complementary to other cosmic probes\footnote{Specifically, while the expansion rate (constrained by geometrical probes such as CMB, BAO, and SNe Ia)  can be consistently explained by different `dark energy' models, they predict measurable differences in the evolution of growth rate with cosmic time.} (Koda et al. 2014).  On small scales,  the random motions of galaxies are determined by the gravitational clustering of galaxies, allowing the study of the mass function of dark matter halos (Sheth 1996).

Classical methods used to estimate peculiar velocities (e.g., Scrimgeour 2016 and references therein) compare the velocity derived from the source redshift with that obtained using the Hubble law combined with an independent distance indicator (such as Tully--Fisher, Faber--Jackson, the Fundamental Plane, or SNe Ia). The main drawback of these methods is the very large intrinsic scatter in distance estimate, which results in an error in the determination of peculiar velocities  (from 5 to 20\% of the Hubble recession velocity, depending on the indicator) that grows linearly with distance. This limits the application of these methods to low redshift values ($z\lesssim 0.05$).  Another technique to measure peculiar velocities, without these limitations in redshift,  is the kinetic Sunyaev--Zeldovich effect that has recently started to be measurable (e.g., Lavaux et al.\ 2013).  Alternatively, the effects of the peculiar velocity field can be studied in an indirect way from the anisotropic pattern of galaxy clustering (redshift-space distortion, Kaiser 1987) through the two-point galaxy correlation function, either focussed on small (e.g., Li et al.\ 2006) or large (e.g., de la Torre et al.\ 2013) scales of galaxy clustering. Finally, the peculiar velocity field can be also studied from numerical simulations, such as the recent high precision $N$-body reconstruction of the Local Universe by Hess \& Kitaura (2016), which has reconciled $\rm \Lambda$CDM with CMB-dipole measurements.

Gravitationally lensed quasars provide a scenario in which the direct measurement of transverse\footnote{Note that the other direct methods measure line-of-sight velocities with subsequent large errors arising from the uncertainties in the subtraction of the Hubble flow.} peculiar velocities, even for $z$ greater than 1, is possible\footnote{Kochanek et al.\ (1996) also proposed to measure the peculiar velocity field  but from a very different ground; the astrometric measurement of proper motions of gravitational lenses.} (Wyithe et al.\ 1999, Gil-Merino et al.\ 2005, Poindexter \& Kochanek 2010, Mediavilla et al.\ 2015).  For an ideal galaxy with a smooth distribution of matter the flux magnification of one lensed image (mean magnification)  is determined by the gravitational field and the lens system geometry. However, real galaxies include stars, which at parsec scales break the smoothness of the mass distribution and of the gravitational field, giving rise to large anomalies in the magnification. Thus, magnification can strongly change in the neighborhood of the lensed image, and, owing to the 
{ effective motion of the quasar source}, the brightness of the image  can experience fluctuations around the mean value (quasar microlensing, Chang \& Refsdal 1979, 1984; Wambsganss 2006). As this variability depends on the unknown spatial distribution of the stars, it is studied as a random process whose properties (in particular the spatial scales of variability) can be modeled. 

The basic idea of the present study is to measure the peculiar velocity of lens galaxies by comparing the fundamental frequency of the temporal variability of microlensed quasar images (inferred from observed light curves) with the fundamental frequency of the spatial variations induced by microlensing in modeled light curves (tracks on magnification maps that simulate microlensing variability). Both quantities should be related  by the relative velocity between the lens galaxy stellar distribution, the source, and the observer. The simplest option to achieve this is to use time-event rate detection methods such as determining the zero-crossing or peak rates of the light curves.

In favorable circumstances, this can be done by counting caustic crossings. In the neighborhood of a lensed source we find regions of more or less gentle magnification gradient and one-dimensional loci with very high magnification (caustic curves). The crossing of a caustic by the lensed source due to its 
{ effective transverse motion} is the most conspicuous event of microlensing. This kind of event can appear, depending on the source size, as a very sharp feature in the lensed image light curve. Thus, for our purposes, caustics can be treated like randomly distributed milestones of known mean separation. The distance traveled by a source is proportional to the number of crossed caustics and, assuming Poissonian statistics, the typical deviation to its square root. 

Thus, we can compare the number of caustic crossings with the mean separation predicted by the models to infer the transverse velocity of the lens galaxy (Wyithe et al.\ 2000a, Gil-Merino et al.\ 2005, Mediavilla et al.\ 2015). One practical drawback of this method is that, given the typical transverse velocities, the mean temporal separation between caustics amounts to several years (see Einstein radius crossing times in Mosquera \& Kochanek 2011)  and we need to count enough crossings to reduce the Poissonian noise. To surmount this problem, we propose to follow an ensemble of (properly selected) gravitational lens systems  to add together a statistically significant number of crossings.

Another problem is that, although caustics are intrinsically very sharp, they could be smeared by the source size. For large sources, the intensity of the caustic crossing can be drastically diminished and perhaps confused with other type of microlensing phenomenology (or two caustics can even be blended into one single event). Thus, the best option to identify and count caustics is to use as small a source as possible. The X-ray emitting region of an AGN would be the optimal choice but, unfortunately, the massive X-ray monitoring of lensed quasars seems unaffordable with present and planned facilities. At other wavelengths (mainly the optical) the microlensing events detected may not always be easily identified as a single caustic crossing. An obvious generalization that includes caustic crossing and other more complex events is to study the rate of Peaks Over a Threshold (POT), which have been considered in the literature mainly for the case of a relatively high threshold (High Magnification Events, HME; e.g., Kayser et al.\ 1986, Kundic \& Wambsganss 1993). Alternatively, we could also study the zero-crossing rate but, then, a no microlensing baseline is needed to set the zero.

The main objetive of this paper is, thus, to propose and discuss the use of microlensing magnification event rates in an ensemble of gravitational lenses to estimate the galaxy peculiar velocities. In \S 2  the measurement of the transverse peculiar velocity from the statistics of microlensing events is introduced. In \S 3 we conduct an illustrative analysis based on POT counting in the  microlensing light curves available in the literature. \S 4 is devoted to analyzing the sources of uncertainty, studying the selection of the ensemble of lens systems, and discussing future perspectives. Finally, the main results are summarized in Section 5.

\section{Methodology \label{apendiceA}}

\subsection{Transverse and effective velocities}

The microlensing event-rate depends on the relative velocity between the source and the spatial pattern of magnifications induced by microlensing, which, in turn, varies with the random proper motion of the stars. The
{ effective transverse velocity of the source}, $\vec v_t$, results from the composition of the relative movements of observer, lens galaxy, and quasar source. Following Kayser et al.\ (1986) the transverse velocity, measured at the source plane using the observer's time, can be written as,

\begin{equation}
\label{kayser}
\vec{v}_t={\vec{v}_o\over 1+z_l}{D_{LS}\over D_{OL}}-{{{\vec{v}}_{\rm pec}(z_l)}\over 1+z_l}{D_{OS}\over D_{OL}}+{{{\vec{v}_{\rm pec}}(z_s)}\over 1+z_s}.
\end{equation}
where ${\vec{v}}_{\rm pec}(z)$ is the peculiar transverse velocity at redshift $z$, $z_l$ and $z_s$ are the redshifts of lens and source, and $\vec{v}_o$ is the observer's velocity, i.e., the projection of the heliocentric CMB dipole velocity onto the lens plane. $D_{OL}$, $D_{OS}$, and $D_{LS}$ are angular diameter distances between observer and lens, observer and source and lens and source, respectively. Defining, for convenience (Kochanek 2004),

\begin{eqnarray}
{\vec{\hat v}_{o}}={\vec{v}_{o}\over 1+z_l}{D_{LS}\over D_{OL}}, \nonumber\\
\vec{\hat v}_{\rm pec}(z_l)={{{\vec{v}}_{\rm pec}(z_l)}\over 1+z_l}{D_{OS}\over D_{OL}}, \nonumber\\
\vec{\hat v}_{pec}(z_s)={{{\vec{v}}_{\rm pec}(z_s)}\over 1+z_s},
\end{eqnarray}
Equation \ref{kayser} can be written as,

\begin{equation}
\label{kayser_short}
\vec{v}_{t}={\vec{\hat v}_{o}}-\vec{\hat v}_{{\rm pec}}(z_l)+\vec{\hat v}_{{\rm pec}}(z_s).
\end{equation}

On the other hand, owing to the random proper motions of the stars in the lens galaxy, the features of the magnification pattern used to define the magnification events (such as peaks or caustics) are not static but move randomly, thereby increasing the average event-rate. According to the simulations by Kundic \& Wambsganss (1993),
the effects of the 
{ effective transverse velocity of the source}, $\vec v_t$, and of the
random stellar kinematics, are about equal when we compare the rms velocity of the stars in the plane of the galaxy perpendicular to the line of sight, $\sqrt{2} \sigma_*$ ($\sigma_*$ is the one-dimensional stellar velocity
dispersion in the lens plane), with the 
{ effective transverse velocity of the source}.
Consequently (Kundic \& Wambsganss 1993), the event-rate would be related to a { composed} effective velocity, $v_{eff}$, 
\begin{equation}
\label{eq0}
{v_{eff}}=\sqrt{v_t^2+a^2\hat{\sigma}_*^2},
\end{equation}
where, $\hat{\sigma}_*={{\sigma}_*\over 1+{z_l}}{{D_{OS}} \over {D_{OL}}} $, { and $a^2$ is an effectiveness parameter. We adopt $a^2= 2$, the expected value when the galaxy bulk motion, $v_t$, and the rms stellar velocity,  $\sqrt{2} \sigma_*$, have equal effects.}\footnote{{ Kundic \& Wambsganss (1993) estimate from simulations $a^2\simeq 1.7$, but }the exact value of the { effectiveness} factor could depend on optical depth and shear (Wyithe et al.\ 2000b). { In any case, reasonable changes in $a^2$ have little impact on our results (see \S \ref{effectiveness}).}}

\subsection{Statistics of microlensing event counts for an ensemble of lensed quasar images}

Let us now consider an ensemble of $M$ gravitationally lensed images, ($i=1,...,M$). The probability of observing, in one image, $n_i$ events (either caustic crossings, zero-crossings, POT, or others) { in a monitoring time $t_i$} conditioned to a given transverse velocity of the lens galaxy, $\vec v_{{t}_i}$, can be formally written as,

\begin{equation}
\label{eq1}
p_i(n_i|v^k_{{t}_i};\sigma_{*_i},\xi_i^j),
\end{equation}
where $v^k_{{t}_i}=(v^1_{{t}_i},v^2_{{t}_i})$ are the components of $\vec v_{{t}_i}$,  $\sigma_{*_i}$ is the 1D stellar velocity dispersion, and $\xi_i^j$ are other physical parameters of interest representative of the lensed image, such as the fraction of mass in microlenses in the lens galaxy, or their mean mass. From microlensing simulations that should include stellar random motions (see Kundic \& Wambsganss 1993 and Poindexter \& Kochanek 2010), we can infer $p_i(n_i|v^k_{{t}_i};\sigma_{*_i},\xi_i^j)$  and from this quantity the likelihood of ${v_{t}}_i$ using Bayes' theorem, 

\begin{equation}
\label{eqlike}
L_i(v^k_{{t}_i};\sigma_{*_i},\xi_i^j|n_i)\propto p_i(n_i|v^k_{{t}_i};\sigma_{*_i},\xiï_i^j).
\end{equation}
Integrating $L_i(v^k_{{t}_i};\sigma_{*_i},\xi_i^j|n_i)$, we can obtain the marginal probabilities, $L_i(v^1_{{t}_i};\sigma_{*_i},\xi_i^j|n_i)$,  and $L_i(v^2_{{t}_i};\sigma_{*_i},\xi_i^j|n_i)$. Finally, using Eq.\ \ref{kayser_short}, we can relate the convolution of the PDFs of the peculiar velocity at the lens and source redshifts with the PDF of the transverse velocity,

\begin{eqnarray}
\label{eqtwins}
L_i(\hat v^1_{pec}(z_l))\ast L_i(-\hat v^1_{\rm pec}(z_s))=L_i(-\hat v^1_{{t}_i}+\hat v^1_{o_i})\nonumber\\
L_i(\hat v^2_{pec}(z_l))\ast L_i(-\hat v^2_{\rm pec}(z_s))=L_i(-\hat v^2_{{t}_i}+\hat v^2_{o_i}).
\end{eqnarray}
These equations can be used in several ways. If the ensemble is large enough, it is possible  in principle to select subsamples in lens redshift and select the systems in which $L_i(-\hat v^{1,2}_{\rm pec}(z_s))$ is presumably narrow to derive from Equations \ref{eqtwins} the average frequency distribution of  peculiar velocities at a given redshift, $\langle L(\hat v^{1,2}_{\rm pec}(z_l))\rangle\sim {1\over M}\sum_i{L_i(-\hat v^{1,2}_{{t}_i}+\hat v^{1,2}_{o_i})}$.

If the limitations of our ensemble do not allow a detailed study of the PDF, we can directly multiply Equation \ref{kayser} by itself, average and use Eq.\ \ref{eq0} to obtain (see Appendix \ref{AA} for the details),

\begin{eqnarray}
\label{eqlong}
{v_{eff}}_{i}=\nonumber \\
\sqrt{\left({v_{oi}\over 1+{z_l}_i}{{D_{LS}}_i\over {D_{OL}}_i}\right)^2+\left({\sqrt{2}{{{\sigma}_{*}}}_i\over 1+{z_l}_i}{{D_{OS}}_i\over {D_{OL}}_i}\right)^2+\left({\sqrt{2}{{\sigma}_{\rm pec}({z_l}_i)}\over 1+{z_l}_i}{{D_{OS}}_i\over {D_{OL}}_i}\right)^2+\left({\sqrt{2}{{\sigma}_{\rm pec}({z_s}_i)}\over 1+{z_s}_i}\right)^2},
\end{eqnarray}
where ${{\sigma}_{pec}({z})}$ is the one-dimensional dispersion of the peculiar velocity field at redshift $z$ (see Appendix \ref{AA}). This equation is in agreement with the expression adopted by Blackburne (2009) and with the formula used by Mosquera \& Kochanek (2011).\footnote{Assuming that Mosquera \& Kochanek consider the transverse (2D) peculiar dispersion velocity instead of the one-dimensional one.}

Following Kochanek (2004; see also Blackburne 2009), in the linear approximation regime of $\Lambda$CDM cosmology it is possible to express ${{\sigma}_{\rm pec}({z})}$  in terms of the cosmological growth factor rate, $f$, and of the peculiar velocity dispersion at zero redshft, ${{\sigma}_{\rm pec}({0})}$,

\begin{equation}
\label{eqs0}
\sigma_{\rm pec}(z)={\sigma_{\rm pec}(0)\over (1 + z)^{1/2}}{f(z)\over f(0)} 
\end{equation}
to write Equation \ref{eqlong} as,

\begin{eqnarray}
\label{eqlong2}
{v_{\rm eff}}_{i}=\sqrt{\left({v_{oi}\over 1+{z_l}_i}{{D_{LS}}_i\over {D_{OL}}_i}\right)^2+\left({\sqrt{2}{{{\sigma}_{*}}}_i\over 1+{z_l}_i}{{D_{OS}}_i\over {D_{OL}}_i}\right)^2}\nonumber \\  
\overline{+\left({\sqrt{2} \sigma_{\rm pec}(0)\over f(0)}\right)^2 \left[\left({f({z_l}_i)\over \left(1+{z_l}_i\right)^{3/2}}{{D_{OS}}_i\over {D_{OL}}_i}\right)^2+\left(f({z_s}_i)\over \left(1+{z_s}_i\right)^{3/2}\right)^2\right]}.
\end{eqnarray}

Using a model for the growth factor rate (see Lahav et al.\ 1991, for instance), it is straightforward to jointly fit all the data from the ensemble of images to measure $\sigma_{pec}(0)$ or, sampled in 2D bins according to $(z_l,z_s)$, to probe the growth factor at different $z$, $f(z)$.

\section{Analysis of microlensing light curves from the literature \label{lc}}

\subsection{Detected POT \label{counts}}

In spite of the heterogeneity (in time sampling and coverage, S/N ratio, or photometric band, for instance) of the available data, we have searched the literature for light curves of gravitational lens systems to identify and  collect POT to compare with the model predictions. Our primary intention is to illustrate and test with real data the method  described above (\S \ref{apendiceA}) to measure transverse velocities. In this process we will obtain some estimates for the transverse velocity, which should nevertheless be considered with caution according to the lack of suitability of the data. 

The first step in the search of microlensing POT is to remove the intrinsic quasar variability, which should be the same in two images of a lensed quasar, but with a time delay caused by the difference between the optical paths of the two images. Thus,  microlensing variability curves free from intrinsic variability are obtained by subtracting the light curves of two images after shifting one of them by the time delay. In most cases  in the literature, microlensing variability curves are directly published by the authors, in other cases we have had the opportunity to compute them using the data made available by the authors, and, finally, in a few cases we have directly compared the published plots of light curves. In a microlensing light curve, $m_A-m_B$, for instance, { we may count both upward and downward peaks that correspond to microlensing in B and A, $n_B$ and $n_A$, respectively. We should, then, compare $n_A$ ($n_B$) with the number of POT in simulated light curves for A (B). In this way, we are using twice the light curve and we need to count twice the monitoring time (but we should avoid more than one duplication if the same image is repeated in different microlensing light curves: A-B, C-B, D-B, for instance).}

Notice that any error in the time delay correction may result in the interpretation of intrinsic variability as microlensing and, hence, these errors (such as fluctuations due to noise) will tend to increase the number of peaks and the estimated velocity. Thus, from the point of view of the event rate counts, the likely effect of errors and uncertainties in the light curves and in time delays is to overestimate the velocity.

To identify POT we start studying every local maximum of intensity $m(x_{max})$. We move from the local maximum towards the left (right) until a minimum in the light curve is reached at $x_{min-}$ ($x_{min+}$) and we take this point to define the experimental baseline by the left (right) of the event. Then we measured the left (right) amplitude of the peak as $\Delta m_{-}=m(x_{max})-m(x_{min-})$ ($\Delta m_{+}=m(x_{max})-m(x_{min+})$) and calculate the S/N ratio from $n_{-}=\Delta m_{-}/\sigma$ ($n_{+}=\Delta m_{+}/\sigma$), where $\sigma$ is an estimate of the light curve noise.

Sometimes, seasonal gaps and other monitoring interruptions can coincide with the peak of the event. When the slopes of the light curves indicate that a maximum occurs within a gap, we estimate a lower limit of the amplitude of the event by the left (right) taken as lower limit for the intensity of the maximum the value of the light curve at the left (right) border of the gap. 

In Table 1 we have included the amplitudes and S/N ratios by the left (right), of the events candidate to POT. When either the minimum or the maximum is not reached owing to a discontinuity, we indicate that the amplitude is a lower limit. When the maximum is lost in a gap we mark the epoch of the peak as approximate. 
Only peaks with amplitude greater than 3$\sigma$ and threshold greater than 0.1 mag qualify as POT. We count nine POT for a threshold of 0.1 mag.  When we rise the threshold to 0.2 mag, we are left with seven POT. { In Table 1 we also give the effective monitoring times. The effective monitoring time corresponds to the time from the first to the last data point of a light curve including seasonal gaps.  Notice that both observed and simulated microlensing events are very much broader than typical seasonal gaps and that we have counted events that peak within a gap.}

\subsection{Modeling POT detection rate}

The likelihood of a given transverse velocity, ${v_{t}}_i$, conditioned to the detection of $n_i$ POT { in a monitoring time $t_i$} can be written as (see Equation \ref{eqlike}),
\begin{equation}
\label{eqlike3}
L_i({v^k_{t}}_i;\sigma_{*_i},R_i,\alpha_i,\langle m_i\rangle|n_i)\propto p_i(n_i|{v^k_{t}}_i;\sigma_{*_i}, R_i,\alpha_i,\langle m_i\rangle), 
\end{equation}
where, in addition to the stellar velocity dispersion $\sigma_{*_i}$, we have made explicit other physical parameters of interest of the lens system, namely, the half-light radius of the source, $R_i$, the fraction of mass in microlenses, $\alpha_i$, and the mean mass of the microlenses, $\langle m_i\rangle$. In principle we could simulate microlensing magnification maps including random proper motions of the stars to calculate $p_i(n_i|{v^k_{t}}_i;\sigma_{*_i}, R_i,\alpha_i,\langle m_i\rangle)$ for each lensed image, and use Equations \ref{eqtwins} to derive information about the PDF of the peculiar velocity field. However, owing to the limitations of our ensemble, we will follow a less ambitious procedure and try to relate directly the effective velocities of the lensed images (see Equations \ref{eqlong} and \ref{eqlong2}) to the total number of POT detected in the ensemble of lensed images.  This work can be made easier if, according to the results by Mediavilla et al.\ (2015) (obtained in the limiting case in which the events are caustic crossings),  we accept that the expected number of POT in a lensed image follows a similar dependence with $\alpha$, $R$, $v_{\rm eff}$, and $\langle m \rangle$,

\begin{equation}
\label{eqext}
\langle n_i \rangle={{v_{\rm eff}}_i t_i \over \sqrt{\langle m \rangle/0.3 M_\odot}}{{l_1}^{-1}_i(R,\alpha)}
\end{equation}
where { $t_i$ is the monitoring time and} ${l_1}_i(R,\alpha)$ is the average track length to obtain a POT for microlenses of mass $0.3 M_\odot$ when the half-light radius of the source is $R$ and the fraction of mass in microlenses is $\alpha$. This equation is formally true for identical mass particles and, after the calculations of Mediavilla et al.\ (2015), it was found to hold also for a comprehensive family of microlens mass distributions. We have checked using numerical simulations that, for the typical case of a lens image with $\kappa=\gamma=0.45$, Eq. \ref{eqext} holds for different values of $\alpha$ and $R$, and that the typical deviation of $\langle n \rangle$ is $\sqrt{\langle n \rangle}$, as expected for a Poissonian variable.

The total number of detected POT is then

\begin{equation}
\label{eqnavg}
\langle n \rangle = \sum_i {\langle n_i \rangle} = \sum_i{{{v_{\rm eff}}_i t_i \over \sqrt{\langle m \rangle/0.3 M_\odot}}{{l_1}^{-1}_i(R,\alpha)}}.
\end{equation}

The next step is to use this last equation (in combination with Equations \ref{eqlong} or \ref{eqlong2}) and the experimental number of POT derived in \S \ref{counts} to illustrate the method, to constrain the peculiar velocity of lens galaxies, and to explore the future possibilities of the method.

\subsection{Results \label{rya}}

\subsubsection{Estimates of $\sigma_{\rm pec}(0.53)$ and $\sigma_{\rm pec}(0)$ \label{sigma0}}

In our ensemble of lensed images the redshift of the lenses is rather concentrated with  $\langle z_l \rangle= 0.53\pm 0.18$. On the other hand, using the $f(z) \approx \left[\Omega_m(z)\right]^{0.6}$ approximation ({ Peebles, 1980}, Lahav et al.\ 1991) and Eq. \ref{eqs0},  we find { (adopting a flat universe with $\Omega_{m_0}=0.317$, $\Omega_{\Lambda_0}=0.683$ and the formulae from Lahav et al. [1991] to compute $\Omega_m(z)$) } that the contribution of the galaxy velocity term in Eq.\  \ref{eqlong} is substantially greater than that of the source (by a factor ranging from 2.23 to 15.5 depending on the system).  Under these favorable circumstances,\footnote{For a larger ensemble, we may have selected the lens systems fulfilling the last condition and falling in a given redshift bin.} we can directly estimate the peculiar velocity at the average lens redshift.  Thus, neglecting $\sigma_{\rm pec}({z_s}_i)$ and approximating, $\sigma_{\rm pec}({z_l}_i)\simeq\sigma_{\rm pec}(\langle z_l \rangle=0.53)$, we can numerically resolve Eqs.\ \ref{eqlong} and \ref{eqnavg} for the $\langle n \rangle=9$ POT counted for a threshold of 0.1 mag and the $t_i$ listed in Table 1. We calculate $v_{oi}$ as the projection of the cosmic microwave background dipole velocity (Hinshaw et al.\ 2009) onto the lens plane. The stellar velocity dispersion, $\sigma_{*i}$, is estimated using the image separation, $\Delta \theta_i$, { given by the SIS model}  (e.g., Treu et al.\ 2009),\footnote{For { two} systems of very large separation, we have used the average value { of the other systems in our sample}, $\langle \sigma_{*}\rangle=260\pm 59\, \rm km\, s^{-1}$.}

\begin{equation}
\Delta \theta_i={8\pi } \left({\sigma_{*i}\over c}\right)^2{D_{LS}\over D_{OS}}.
\end{equation}
To calculate the factor ${{l_1}^{-1}_i(R,\alpha)}$ in Eq.\ \ref{eqnavg}, we compute, using the Inverse Polygon Method (Mediavilla et al.\ 2006, Mediavilla et al.\ 2011), microlensing magnification maps for each lensed image (characterized by $\kappa$ and $\gamma$) for different values of $\alpha$. {Each map is posteriorly convolved with { the Sakhura \& Sunyaev (1973) disk model for different half-light radii, $R$ (we have considered also sources with a Gaussian luminosity profile for the discussion in \S \ref{profile})}. Finally, we obtain a large number of tracks (10\,000) from each map and count, following the same prescription used in the experimental light curves, POT for a threshold of 0.1 mag to evaluate ${{l_1}^{-1}_i(R,\alpha)}$. The resulting $\sigma_{\rm pec}(0.53)$ obtained solving Eq.\ \ref{eqnavg} and Eq. \ref{eqlong} numerically is plotted in Figure 1 as a function of $R$ and $\alpha$.

Following the same steps to infer $v_{oi}$, $\sigma_{*i}$, and ${{l_1}^{-1}_i(R,\alpha)}$, but now without neglecting the source velocity term,  we can solve Eqs.\ \ref{eqlong2} and \ref{eqnavg} numerically to obtain $\sigma_{\rm pec}(0)$ as a function of  $R$ and $\alpha$ (see Figure 2).

Figures 1 and 2 show that, as expected, the estimated peculiar velocities increase with $R$. On the contrary, the dependence with $\alpha$ is less important in the region of interest (see below). Several authors have studied the size of  quasar accretion disks from the analysis of microlensing magnification. Results by Mosquera et al.\ 2013 (see also Morgan et al.\ 2010) confirm the $R\propto M^{2/3}_{BH}$ dependence of the size with the central black-hole mass predicted by thin disk theory. The average black-hole mass (weighted according to the monitoring time)  of the systems in our ensemble with available determinations (Peng et al.\ 2006, Assef et al.\ 2011) is $\langle M_{\rm BH}\rangle=0.68 \times 10^9 M_\odot$, which according to Mosquera et al.\ (2013) corresponds to a size of 3.6 light-days. From the statistics of microlensing magnification amplitudes, Jim\'enez-Vicente et al.\ (2015a,b) have estimated a value of $R=7.9^{+3.8}_{-2.6}$ light-days for an average $1.36 \times 10^9 M_\odot$ mass black-hole and $\alpha=0.21\pm 0.05$. Normalizing to { the $0.68 \times 10^9 M_\odot$ average black-hole} mass using the $R\propto M^{2/3}_{\rm BH}$ dependence, we obtain a size of $5.0^{+2.4}_{-1.6}$ light-days in reasonable agreement with the results based in Mosquera et al. { (2013)}. If we take the average between both measurements, $R=4.3$ light-days, as estimate for the average size of our ensemble, we obtain (see Figure 1), $\sigma_{\rm pec}(0.53)\simeq (638\pm 213) \sqrt{\langle m \rangle/0.3 M_\odot} \, \rm km\, s^{-1}$. The error here corresponds to Poissonian noise in the POT count.  This measurement  is independent of the cosmological model (except in the computation of angular distances)  and, to our knowledge, is the highest direct redshift determination of $\sigma_{\rm pec}$. For the reference value, $R=4.3$ light-days, we can also estimate the peculiar velocity at zero redshift (see Figure 2), $\sigma_{\rm pec}(0)\simeq (491\pm 164) \sqrt{\langle m \rangle/0.3 M_\odot} \, \rm km\, s^{-1}$.

\subsubsection{Comparison with other results and with $\Lambda$CDM predictions \label{comp}}

It is interesting, even within the limitations imposed by the shortcomings of the available data, to compare our estimates with other results and model predictions. {  In our neighborhood, peculiar velocities are measured using secondary distance indicators (such as Tully--Fisher, Faber--Jackson, the Fundamental Plane, or SNe Ia).   These studies measure the line of sight component of the peculiar velocity field, $S_n$, for a survey of objects. The problem is that the error of each radial velocity estimate, $\sigma_n$, is very large (from 5\% to 20\% of the Hubble recession velocity depending on the distance indicator, typically,  $\sigma_n> 10^3\,\rm km\, s^{-1}$) and, consequently, the distribution of peculiar velocities is substantially broadened by the errors. Aware of this problem, Tully et al. (2013) have obtained the histogram of peculiar velocities for single galaxies and groups with fractional distance error estimates $\lesssim 0.14$ and a histogram from mock galaxies  with distances randomly assigned taking into account the error for each case. From these histograms we derive $\sigma_{pec} \simeq 426\pm 85\, \rm km\, s^{-1}$, in agreement to within the uncertainties with our estimate (see Figure  \ref{sig}).  

{ Many peculiar velocity studies suppose that the $S_n$ are drawn from a Gaussian distribution of mean $\vec v$, the bulk flow, and with variance $\sigma_n^2+\sigma_{**}^2$, where  $\sigma_{**}$ represents the contribution of small spatial scale fluctuations.\footnote{{ This decomposition implies that $\sigma_{**}\lesssim \sigma_{pec}$ as far as peculiar motions of lens galaxies are affected by the distribution of matter on both large  (dominated by cosmological overdensities) and small (influenced by dark matter halos) scales.}} We find our estimate for $\sigma_{pec}$ consistent with $\sigma_{**}$ determinations available in the literature (see Figure \ref{sig})}.}
{{ The other outcome of peculiar velocity studies is the bulk flow. It is obtained}  averaging the peculiar velocity field of a certain volume of universe centered on us, integrating down  the contribution from small spatial scale fluctuations. Thus, bulk flows represent only a lower limit to the peculiar velocity, and any comparison with our data should be made for the smaller integration windows in redshift.  The bulk flow determined for the smallest region is the Local Group (LG) speed, $|\vec v_{\rm LG}|=627\pm 22\, \rm km\, s^{-1}$ (Kogut et al.\ 1993). Accepting a Maxwell--Botzmann distribution for the peculiar velocity field { (a simplifying assumption; see however Scrimgeour et al. 2016 and references therein)}, we obtain, $\sigma^{\rm LG}_{\rm pec}= |\vec v_{\rm LG}|/\sqrt{8/\pi}=393\pm 14\, \rm km\, s^{-1}$, in agreement with our estimate (see Figure  \ref{sig}). 

Linear perturbation $\Lambda$CDM models based on WMAP9 cosmology (Carrick et al.\ 2015) predict results also compatible within errors with our estimate\footnote{We could have made the comparison in the opposite way by computing from our estimate of $\sigma_{\rm pec}(0)$ the amplitude of the peculiar velocity at zero redshift, $|\vec v_{\rm pec}(0)|\simeq (784\pm 261) \sqrt{\langle m \rangle/0.3 M_\odot} \, \rm km\, s^{-1}$.} (see Figure \ref{sig}), but some way in tension with the better constrained amplitude of the LG motion, which, it has been argued, may be a fluctuation due to cosmic variance. On the contrary, recent results by Hess \& Kitaura (2016), using non-linear phase-space reconstructions of the Local Universe, find the LG speed compatible with $\Lambda$CDM models. In fact, their predictions for LG-like haloes contained within a distance of $2.3\, h^{-1}\,\rm Mpc$ from the observer imply $\sigma^{\rm LG}_{\rm pec}= |\vec v_{\rm LG}|/\sqrt{8/\pi}=439\pm 69\, \rm km\, s^{-1}$ (see Figure  \ref{sig}), in excellent agreement with the experimental $\sigma^{\rm LG}_{\rm pec}$. 

Looking at Figure  \ref{sig}, we can conclude { that $\sigma_{\rm pec}$ estimates} are better explained by the nonlinear $\Lambda$CDM model.  This is reasonable in so far as the motions associated with small scale galaxy clustering are also expected to contribute significantly to the velocity dispersion of galaxies. This contribution can be studied from the peculiar pairwise velocity dispersion through the two-point galaxy correlation function.  According to Li et al.\ (2006), the dispersion of the one-dimensional peculiar velocities can be in the range of 400 to 800$\, \rm km\, s^{-1}$. However, these values depend on the types of galaxies, and a comparison with the peculiar velocities of lens galaxies should be made with caution. Ultimately, the small scale contribution to the peculiar velocity field may depend on the location of the galaxy in a dark matter halo and be significantly different if the galaxy is a satellite or the central host (Tinker et al.\ 2007). It is beyond the scope of the present work to study this question but the expected contribution of small scale clustering to the peculiar velocity of lens galaxies may be significantly smaller than the pairwise velocities obtained for typical combinations of host and satellite galaxies.

We have also included in Figure  \ref{sig} the individual measurement of the peculiar velocity made from caustic crossing counts in Q2237+0305, a lens system with the galaxy at an unusually low redshift, $z_l=0.04$. The peculiar velocity determination for the lens galaxy (Mediavilla et al.\ 2015), $\sigma_{\rm pec} = (399\pm 231)\sqrt{\langle m \rangle/0.3 M_\odot} \, \rm km\, s^{-1}$, is consistent with the estimate obtained here for $\sigma_{\rm pec}(0)$. 

In summary, our estimate of the peculiar velocity at zero redshift is compatible with { peculiar velocity surveys and} LG kinematics. It is also in agreement to within the uncertainties with linear and nonlinear $\Lambda$CDM models according to recent simulations by Hess \& Kitaura (2016). In principle, the random local velocity field may also contribute substantially to the velocities obtained from microlensing variability. However, the fraction of halo satellite galaxies among lens galaxies can be substantially lower than the typical (Treu et al.\ 2009), so that the peculiar velocities determined using microlensing variability would be related mainly to large scale structure. The study of a larger ensemble of lens systems, with the subsequent reduction in the uncertainties, would allow us to measure the degree of nonlinearity of the galaxies' peculiar motion with regard to the $\Lambda$CDM framework.
}

\section{Discussion}

The primary objective of the study of the light curves of lensed quasars available in the literature carried out in the previous sections was to illustrate and analyze the method of estimating peculiar velocities from microlensing variability. Beyond this initial purpose, the outcomes have proven to be worthy of attention with regard to the study of peculiar velocities at relatively high redshift, although we should bear in mind that they are based on heterogeneous and  irregularly sampled data.

In the coming years new facilities will greatly improve the number and quality of the photometric observations of gravitational lenses. The Large Synoptic Survey Telescope (LSST), for instance (Marshall et al.\ 2010), will discover and monitor around 2600 new systems of lensed quasars, of which $\sim$14\% will be quads. This implies around 6000 lensed images potentially available to search for microlensing events. The lens galaxies will be distributed from $z\sim 0$ to $2$. A cadence of the photometric monitoring of about one measurement per week, will be sufficient to sample microlensing event peaks (and to start alerts for better sampled follow-up monitoring when needed).

In the following paragraphs we are going to discuss the future perspectives of the microlensing-based study of peculiar velocities. Specifically,  we will analyze the sources of uncertainty,  error mitigation strategies, and possible outcomes  for large surveys such as the LSST, taking as reference our study of the 17 lens system ensemble (\S \ref{lc}). 

\subsection{Sources of uncertainty}

\subsubsection{Uncertainties in the experimental count of POT. Poissonian noise}
The error in the experimental detection of the microlensing peaks is a source of uncertainty that can be controlled using the threshold. With this aim, we have repeated all the previous computations of \S \ref{lc} but considering now the seven POT detected with a threshold of 0.2 mag. The results, $\sigma_{\rm pec}(0.53)\simeq (657\pm 248) \sqrt{\langle m \rangle/0.3 M_\odot} \, \rm km\, s^{-1}$ and $\sigma_{\rm pec}(0)\simeq (506\pm 191) \sqrt{\langle m \rangle/0.3 M_\odot} \, \rm km\, s^{-1}$ agree within errors with the estimates derived using the 0.1 mag threshold.  This agreement is important because it supports both the consistency of the method and the limited impact on the results of the errors in the process of POT identification  from the experimental light curves. We have adopted the results for the 0.1 mag threshold (considering that the 0.2 mag threshold would diminish the influence of the photometric errors but at the cost of decreasing the number of POT and, hence, incrementing the Poissonian noise). In the case of the large ensemble of lenses expected from the LSST survey, affected by a very small Poissonian error, the threshold could be significantly increased, making irrelevant the error in the POT count.

We have checked the impact of maxima within gaps with amplitudes between 0.1 and 0.2 mag, which we have considered as POT only for the 0.1 threshold but which may also qualify for the 0.2 threshold. We have found only two cases which would at most increase  the velocity determined using the 0.2 mag threshold by a factor $\sim 2/7$, below the Poissonian error ($\sqrt7/7$). There are also three maxima in gaps not qualified as POT which could imply, in the most conservative case, an increase in velocity comparable to the Poissonian error.

According to Table 2, Poissonian noise is the dominant source of error in the  17 lenses study, based on the statistics of the sum of all the events detected in the ensemble. Although in a more ambitious study (LSST survey) the methodology might be better based in the determination of the PDF (Eqs.\ \ref{eqtwins}), it is practical to take the Poissonian noise as a reference also in this case.  Thus, if for the LSST survey we consider ensembles of 500 images to subsample in redshift, we will reach a Poissonian noise of $\sim 0.07$ (extrapolating the number of counts measured in a total of $\sim 300$ years for the available ensemble of 17 lensed quasar systems, to 500 images monitored during 10 years). If we repeat the same calculation with all 6000 images (that can be used together if a functional dependence of the growth factor rate with redshift, e.g. $f \approx \Omega_m^{0.6}$, is assumed) the Poissonian error in the number of counts will be $\sim 0.02$.

\subsubsection{Uncertainties in source size, $R$, fraction of mass in microlenses, $\alpha$, and lens model ($\kappa$, $\gamma$)}

In \S \ref{rya} we saw that there is a dependence of the transverse velocity estimate with the size of the source and, more weakly, with the fraction of mass in microlenses. According to Jim\'enez-Vicente et al.\ (2015a,b) we can expect a dispersion in the size of a lensed quasar source of about 40\%  with respect to the average. This (see Figs 1 and 2) would imply an uncertainty of less than 10\% in the estimate of peculiar velocities. In the future, the uncertainty in $R$ can be mitigated by determining the size of the source for each lens system (e.g., from the size versus supermassive black-hole mass relationship by Morgan et al.\ 2010).  We can calculate, conservatively, a 20\% size error using this relationship  that will result in a difference in peculiar velocity for a given system of less than 5\%. On the other hand, we expect the (non-intrinsic) dispersion in the size vs.\ mass relationship to be reduced in the future. 

There is now much evidence (e.g., Jim\'enez-Vicente et al.\ 2015a,b, Schechter et al.\ 2014 and references therein) supporting the hypothesis that the fraction of mass in microlenses is in the $0.1 \le \alpha \le 0.3$ range. From the statistics of microlensing magnification amplitudes, Jim\'enez-Vicente et al.\ (2015a,b) have estimated a value of $\alpha=0.21\pm 0.05$.  This  error would imply an uncertainty of less than 4\% in the estimate of peculiar velocities (see Figs.\ 1 and 2). On the other hand, recent studies (Oguri et al.\ 2014, Schechter et al.\ 2014, Jim\'enez-Vicente, 2015a,b) show that it is possible to make a reasonable estimate of the radial distribution of dark matter that could help to determine with more precision the fraction of mass in microlenses for each lensed image and reduce the impact of errors in $\alpha$ in the computation of peculiar transverse velocities.

The values of the convergence, $\kappa$, and shear, $\gamma$, at the location of each image are needed to compute the magnification map from which the POT rate is modeled. These values are inferred from the macro-lens model and could be affected by uncertainties. The relationship of the errors in $\kappa$ and $\gamma$ with the uncertainty in the POT rate is not simple and will probably depend on the change in  the magnification of the source. To evaluate the impact that changes in $\kappa$ and $\gamma$ for each image can have on the results of our 17-lens analysis, we have bootstrapped the ($\kappa,\gamma$) pairs within our ensemble. Considering 1000 bootstrap samples we obtain a mean difference of 3\% in the estimate of the effective velocity. Notice that the real impact of changes will be substantially smaller as bootstraping is an easy to apply, but very conservative, way to test changes in the lens model. On the other hand, the uncertainty in $\kappa$ and $\gamma$ may be controlled by excluding from the ensemble systems with high uncertainties in the fitted model and, if this do not imply any bias, systems in which small changes in the parameters imply large changes in the POT rate. 

In any case, the uncertainties in these parameters ($R$, $\alpha$, $\kappa$, and $\gamma$), probably dominated by random errors in their estimate for each lensed quasar, can be greatly diminished by averaging within a large ensemble of lenses (as explicitly seen by bootstrapping in ($\kappa,\gamma$) pairs).  In the case of the planned LSST monitoring,  we will be considering the sum of the events of a great number of systems with similar weight (similar monitoring time), and the uncertainties in the lens models will be even more diluted provided that the ensemble is a reasonably representative sample of the universe of gravitational lens systems.\footnote{In fact, in this case resampling without replacement among the models will have no impact at all.} For this reason, the values corresponding to these parameters in the error budget (Table 2) are probably conservative upper limits.

\subsubsection{Uncertainties in the mean mass of the stellar PDMF, $\langle m\rangle$, and in the luminosity profile of the source \label{profile}}

Unlike the previous parameters, these sources of uncertainty may induce systematic errors in the estimate of peculiar velocities that could not be averaged out within the ensemble. The shape of the source luminosity profile may have an impact on the number of counts, at least when a profile with a rather smooth core, such as a Gaussian, is compared with the relatively cuspy core profile of the Shakura \& Sunyaev (1973) thin disk model.  According to simulations not shown here, for typically expected sizes of $R\sim4$ light-days, a difference in peculiar velocity estimate of about 7\% is possible.  Thus, there exists a sensitivity of the rate of POT to the shape of the profile of the accretion disk that may be significant for large ensembles with small Poissonian noise. Although the Shakura \& Sunyaev (1973) model seems to be a reasonable representation of the quasar accretion disk (Mediavilla et al.\ 2015) the future monitoring  with the LSST will greatly improve and generalize the study of the luminosity profile of these objects.

On the other hand, it is important to notice that all velocity estimates are affected by the uncertainties in the mean mass of the microlenses, i.e., by lack of knowledge of the stellar Present Day Mass Function (PDMF) in the lens galaxies.  Equation \ref{eqext} indicates that the dependence on the mass function can be expressed in an explicit way in terms of the mean mass of the microlenses, $v_{\rm pec} \propto \sqrt{\langle m \rangle}$.  This dependence can have a noticeable impact on the determinations of the peculiar velocity, as an uncertainty in the mean mass of the stellar PDMF of 30\% would imply an uncertainty in the velocity of 15\%.  This impact can be controlled by marginalizing on $\sqrt{\langle m \rangle}$ using as prior the updated available information about the PDFM.

\subsubsection{Uncertainties in $\sigma_*$ and $v_o$ \label{effectiveness}}

If the peculiar velocity dispersions are, as we have tentatively measured, relatively high, $\sigma_*$ and $v_o$ can be relevant only for investigating the low velocity range of the PDF of peculiar velocities. However, if they were comparable we should obtain good estimates for $\sigma_*$ from follow-up spectroscopic mesurements. According to Treu et al.\ (2009) errors in $\sigma_*$ are typically $\sigma_{\sigma_*}\lesssim 20 \, \rm km\, s^{-1}$, which should be compared in quadrature with the effective transverse velocity, which is expected to be very much greater. 

{ On the other hand, $\sigma_{pec}^2>>\sigma_*^2$, also implies that, even assuming a very broad variation range for the effectiveness parameter (Eq. \ref{eq0}) between $a^2=1$ and $a^2=3$, the impact  in the estimate of the peculiar velocity dispersion would be of less than a 4\%, well below the Poissonian uncertainty.}

\subsubsection{Error budget}

In Table 2 we summarize the contributions of the different sources of uncertainty to the total error. In the case of the available ensemble of 17 lenses, the sum in quadrature of all the errors, including the Poisonian one, amounts to 35\%,  close to the Poissonian error.  This supports the robustness of the results obtained for even an ensemble of relatively small size. 

For a large ensemble of gravitationally lensed quasars, such as the LSST survey, Poissonian noise would be substantially reduced, and the main contribution to the total error may be from two systematic sources of error that cannot be averaged out over the sample: the shape of the source profile and the mean mass of the microlenses. This last parameter would need to be known with an accuracy better than 15\% to achieve a total error of less than 10\% in the estimate of the peculiar transverse velocities.

\subsection{ Outcome from a large survey. Selection of the ensemble of lens systems}

The analysis of \S \ref{lc} has been focussed on the study of the expected number of counts of the ensemble and could be applied with the consequent reduction in Poissonian noise to a more extended ensemble. However, in a future investigation, based on high quality data of a large ensemble of lens systems, a more ambitious analysis to derive the maximum possible information (the histogram of peculiar velocities at different redshifts ideally) should be attempted. The outcome of this study will be conditioned by the selection of lens systems and subsamples attending to the redshifts of lens and source and the relevance of the peculiar velocity of the source, $\hat \sigma_{\rm pec}(z_s)$. 

If we do not apply any restriction on the potentially useful 6000 lensed images to be monitored by the LSST, we can use them together in combination with Eq.\ \ref{eqlong2} to check the cosmological model (performing, for instance,  a fit to $\sigma_{\rm pec}(0)$ and to the exponent, $x$, of the growth rate factor dependence, $f \approx \Omega_m^x$).  

If we select only lens systems in which $\hat \sigma_{\rm pec}(z_l) >> \hat \sigma_{\rm pec}(z_s)$, then we can estimate from Equation \ref{eqlong2} the histograms and expected values of $\sigma_{\rm pec}(z_l)$ for several bins in redshift, from which the direct computation of the growth factor rate at different redshifts is immediate, $f(z_l)/f(0)=(1+z_l)^{1/2}\sigma_{\rm pec}(z_l)/\sigma_{\rm pec}(0)$. To estimate the number of lens systems fulfilling the $\hat \sigma_{\rm pec}(z_l) >> \hat \sigma_{\rm pec}(z_s)$ condition, we have represented in Figure 3 the $(z_l,z_s)$ pairs of the representative sample of lens systems by Mosquera \& Kochanek (2011) and the curves, $\hat \sigma_{\rm pec}(z_s)/\hat \sigma_{\rm pec}(z_l) =0.05,0.1$. Points above the curves correspond to lens systems that fulfill the condition of having less than 5\% or 10\% respectively of the contribution from the source peculiar velocity. Thus, according to Figure 3, we have wide range of possibilities for selecting systems in which ${\sigma}_{\rm pec}(z_s)$ can be neglected that are useful to directly studying the dependence of the growth factor with redshift, $f(z)$.

On the same assumption, $\hat \sigma_{\rm pec}(z_l) >> \hat \sigma_{\rm pec}(z_s)$, we can go a step forward using Equations \ref{eqtwins} to infer the average PDF of $v^{1,2}_{\rm pec}(z_l)$ in different redshift bins, $\langle L(\hat v^{1,2}_{\rm pec}(z_l))\rangle$, as a useful check for the hypothesis of Gaussianity.

\section{Conclusions}

We have proposed and developed a new method based on the count of microlensing events (peaks over a threshold) in an ensemble of lensed quasar images to infer the transverse velocity of lens galaxies and to study the peculiar velocity field. The main results are the following:

1 - To illustrate the method with a pilot study, we have considered an ensemble of 17 gravitationally lensed quasars with light curves available from the literature, to show explicitly that, in spite of the hetereogeneous quality of the data (in S/N ratio, time coverage, sampling, etc.), the application of the method is straightforward and consistent even when two different thresholds of 0.1 and 0.2 mag are alternatively considered to qualify the peaks.

2 - We have studied the impact of the uncertainties on the parameters involved in the modeling of the microlensing variability rate. We found that most of them have a weak impact and can be, in most cases, averaged out over the ensemble and/or controlled with a suitable selection of lens systems. The most important source of uncertainty is the mean mass of the stellar PDMF, which can nevertheless be expressed in an explicit way in the results (allowing itself to be easily marginalized with the best available information about the PDMF).

3 - Even with the obvious limitations of the data on which they are based, the results deserve attention. We found a tentative estimate of the peculiar velocity dispersion at $z\sim 0.5$, $\sigma_{\rm pec}(0.53)\simeq (638\pm 213) \sqrt{\langle m \rangle/0.3 M_\odot} \, \rm km\, s^{-1}$, independent of the cosmological model (except for the computation of angular distances)  and, to our knowledge, the highest direct redshift determination of the peculiar velocity dispersion.

4 - Using $f \approx \Omega_m^{0.6}$ ({ Peebles 1980}, Lahav et al.\ 1991) to transform the velocity of each lens system to $z=0$, we have also obtained an estimate of  the 1D peculiar velocity dispersion at zero redshift, $\sigma_{\rm pec}(0)\simeq (491\pm 164) \sqrt{\langle m \rangle/0.3 M_\odot} \, \rm km\, s^{-1}$. This result is compatible, to within the uncertainties, with { the results of local velocity surveys}, predictions of recent nonlinear $N$-body models of the Local Universe based on the standard $\Lambda$CDM cosmology and with the Local Group bulk flow.

5 - With the ensemble of 6000 monitored lensed images provided by the LSST, it will be possible to determine the growth factor rate dependence with redshift in the $z\sim 0$ to 2 range. With an adequate selection of lens systems, it would be possible to study the PDF of the peculiar velocity field.

\appendix 

\section{Derivation of the equation relating the second order moments of the transverse velocities \label{AA}}

This derivation follows Kochanek (2004) very closely.  In the first place, we combine in Eq.\ \ref{kayser} the peculiar velocities,
\begin{equation}
\label{combine}
\vec{\hat v}_{\rm pec}(z_l,z_s)=-{{{\vec{v}}_{\rm pec}(z_l)}\over 1+z_l}{D_{OS}\over D_{OL}}+{{{\vec{v}_{\rm pec}}(z_s)}\over 1+z_s},
\end{equation}
and, for each image, define, 

\begin{equation}
\label{hat}
{\vec{\hat v}_{o_i}}={\vec{v}_{o_i}\over 1+z_l}{D_{LS}\over D_{OL}},
\end{equation}
to write Equation \ref{kayser} as

\begin{equation}
\label{a3}
\vec{v}_{t_i}={\vec{\hat v}_{o_i}}+\vec{\hat v}_{{\rm pec}_i}(z_l,z_s).
\end{equation}
For transverse peculiar velocities normally distributed,

\begin{equation}
\label{peculiar1d}
\vec{ v}_{\rm pec}(z)\sim (N(0,\sigma_{\rm pec}^2(z)),N(0,\sigma_{\rm pec}^2(z)),
\end{equation}
we will define an effective peculiar dispersion velocity, $\hat \sigma_{\rm pec}(z_l,z_s)$, from

\begin{equation}
\label{sigmapec}
\vec{\hat v}_{\rm pec}(z_l,z_s)\sim (N(0,\hat \sigma_{\rm pec}^2(z_l,z_s)),N(0,\hat \sigma_{\rm pec}^2(z_l,z_s)),
\end{equation}
where

\begin{equation}
\label{a6}
\hat \sigma_{\rm pec}^2(z_l,z_s)=\left({{{\sigma}_{\rm pec}(z_l)}\over 1+z_l}{D_{OS}\over D_{OL}}\right)^2+\left({{{\sigma}_{\rm pec}(z_s)}\over 1+z_s}\right)^2.
\end{equation}
Thus, from Equations \ref{a3} and \ref{a6}, we have

\begin{equation}
\label{pecB}
\vec v_{ti}\sim (N(\hat v_{oi}^1,\hat \sigma_{\rm pec}^2(z_l,z_s)),N(\hat v_{oi}^2,\hat \sigma_{\rm pec}^2(z_l,z_s))).
\end{equation}

The probability distribution of the length of a vector that has components that are Gaussian-distributed, and that is not centered at zero, is the Rice distribution (compare with Kochanek 2004),

\begin{equation}
\label{transverse}
p(v_{ti})={v_{ti} \over \hat \sigma_{\rm pec}^2}I_0\left[{v_{ti} {\hat v}_{oi}\over \hat \sigma_{\rm pec}^2}\right]\exp\left(-{v_{ti}^2+ {\hat v}_{oi}^2 \over 2\hat \sigma_{\rm pec}^2}\right),
\end{equation}
where $I_0$ is the modified Bessel function of the first kind with order zero. The second moment of this distribution is

\begin{equation}
\label{moment2}
\left\langle{v}_{ti}^2\right\rangle=\hat v_{oi}^2+2{{\hat \sigma}^2_{{\rm pec}}(z_l,z_s)}
\end{equation}
The same result is obtained multiplying Equation \ref{kayser_short} by itself, supposing that the peculiar velocities are normally distributed (Eq.\ \ref{sigmapec}), averaging (${\hat v}_{\rm pec}$ follows a Rayleigh distribution) and taking into account that, by symmetry, $\left\langle\vec{\hat v}_{\rm pec}\cdot \vec{\hat v}_{oi}\right\rangle=0$. Substituting Equation \ref{a6} in Equation \ref{moment2}, we obtain

\begin{equation}
\label{eq14}
\left\langle{v}_{ti}^2\right\rangle=\left({v_{oi}\over 1+z_l}{D_{LS}\over D_{OL}}\right)^2+\left({\sqrt{2}{{\sigma}_{\rm pec}(z_l)}\over 1+z_l}{D_{OS}\over D_{OL}}\right)^2+\left({\sqrt{2}{{\sigma}_{\rm pec}(z_s)}\over 1+z_s}\right)^2.
\end{equation}
As was noted by Kochanek (2004), the same result can be reached by supposing that all variables are Gaussian and can be summed in quadrature.\footnote{Equation \ref{eq14} can be directly obtained multiplying Equation \ref{kayser} by itself, supposing that the absolute value of the transverse peculiar velocity is normally distributed, ${v}_{\rm pec}(z)\sim N(0,2\sigma_{\rm pec}^2(z))$ ($\sigma_{\rm pec}(z)$ is, then, the effective one-dimensional dispersion at redshift $z$), and averaging.} Combining Equations \ref{eq14} and \ref{eq0}, we obtain Equation \ref{eqlong2}.

\acknowledgements{We thank the referee for the thorough review of the paper and for the many useful suggestions. This research was supported by the Spanish MINECO with the grants AYA2013-47744-C3-3-P and AYA2013-47744-C3-1-P. JAM is also supported by the Generalitat Valenciana with the grant PROMETEO/2014/60. JJV is supported by the project AYA2014-53506-P financed by the Spanish Ministerio de Econom\'\i a y Competividad and by the Fondo Europeo de Desarrollo Regional (FEDER), and by project FQM-108 financed by Junta de Andaluc\'\i a.}

\begin{figure}[h]
\plotone{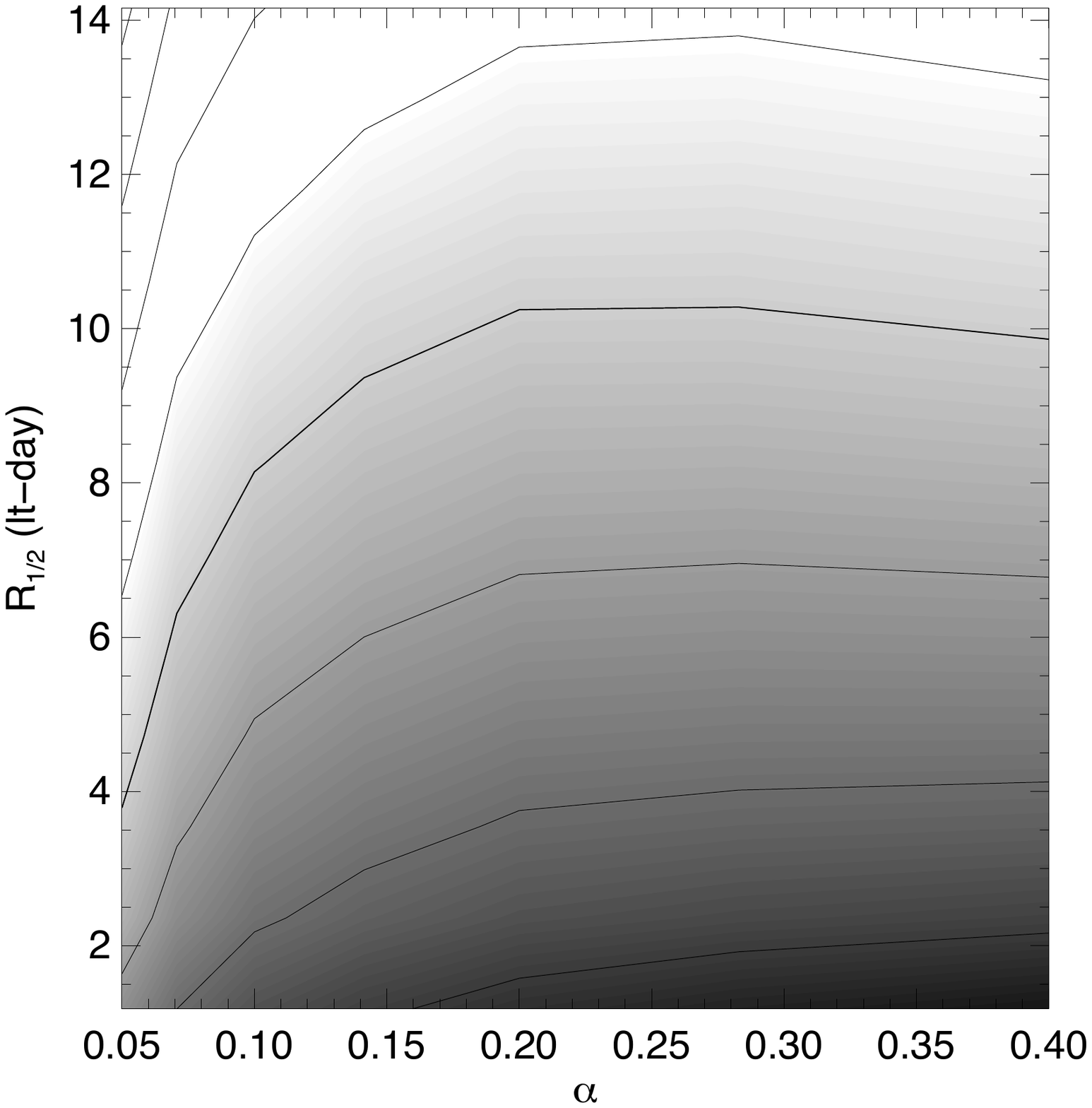}
\vskip -3 truecm
\caption{Estimate of the peculiar velocity dispersion for $z=0.53$ , $\sigma_{\rm pec}(0.53)$, as a function of the fraction of mass in stars in the lens galaxy, $\alpha$, and of the half-light radius of the source, $R$. Lighter gray levels correspond to higher velocities. The contours  range from 400 to 1800$\,\rm km\,s^{-1}$ in steps of 200$\,\rm km\,s^{-1}$. The thicker contour corresponds to 1000$\,\rm km\,s^{-1}$.}
\end{figure}

\begin{figure}[h]
\plotone{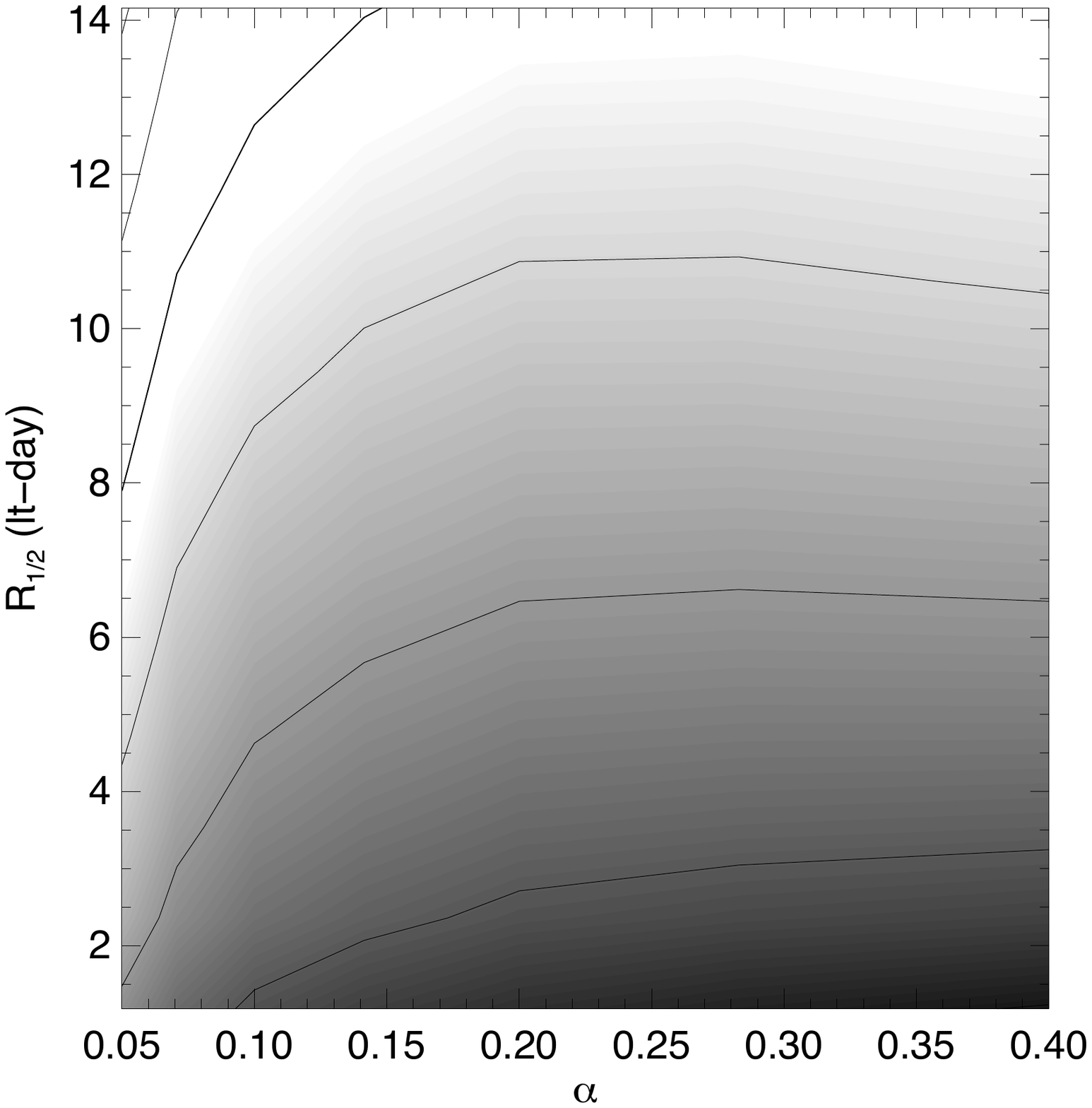}
\vskip -3 truecm
\caption{Estimate of the peculiar velocity dispersion for $z=0$, $\sigma_{\rm pec}(0)$, as a function of the fraction of mass in stars in the lens galaxy, $\alpha$, and of the half-light radius of the source, $R$. Lighter gray levels correspond to higher velocities. The contours  range from 400 to 1400$\,\rm km\,s^{-1}$ in steps of 200$\,\rm km\,s^{-1}$. The thicker contour corresponds to 1000$\,\rm km\,s^{-1}$.}
\end{figure}

\begin{figure}[h]
\vskip -2.5 truecm
\plotone{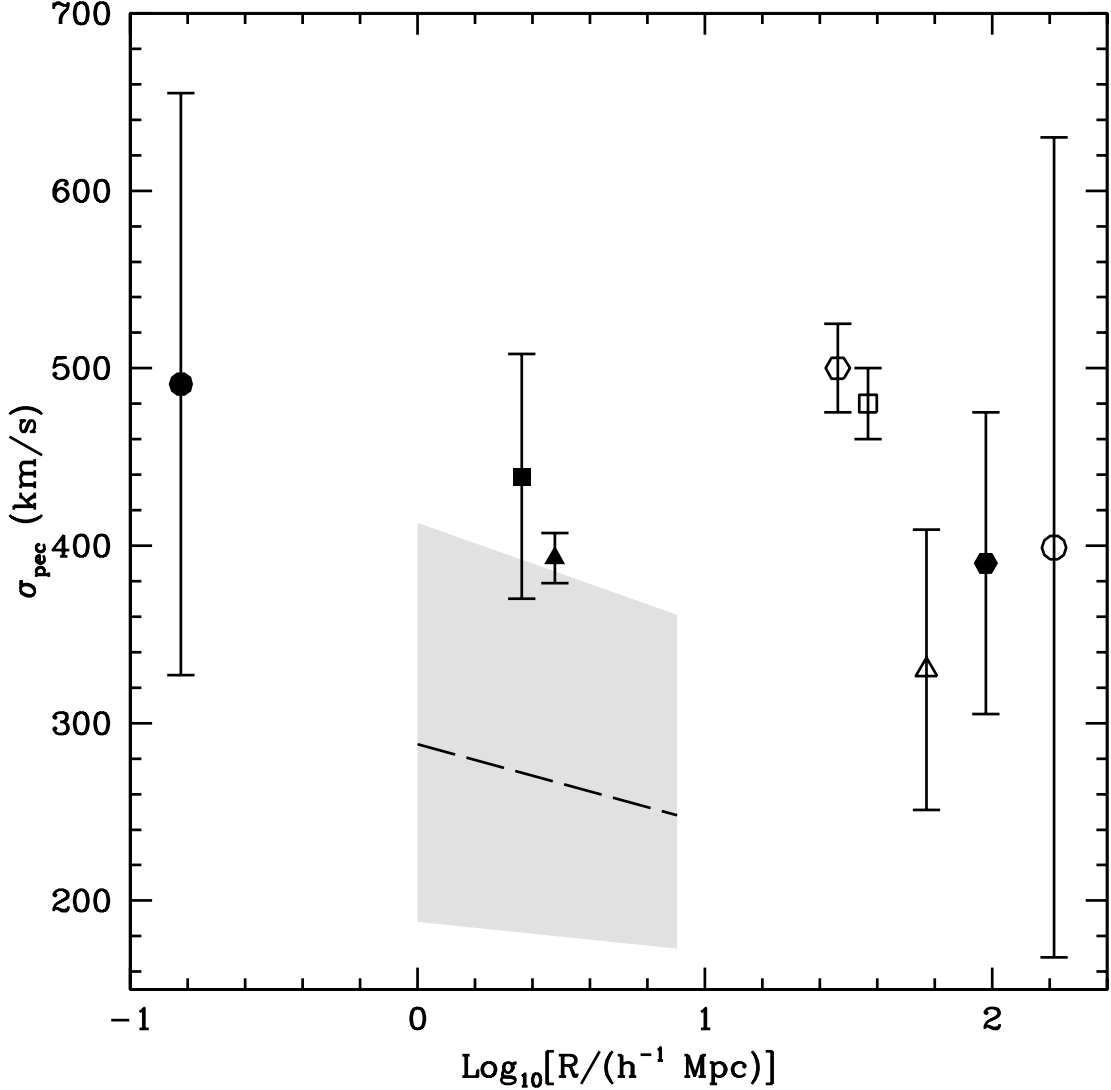}
\vskip -4 truecm
\caption{\label{sig} Peculiar velocity dispersion, $\sigma_{\rm pec}$, versus { characteristic survey depth, $R$} ({ notice that the window functions used to define $R$ differ for each survey}). The filled circle { (arbitrarily located in $R$)} shows the peculiar velocity dispersion at $z=0$, $\sigma_{\rm pec}(0)$, estimated from microlensing variability (this work). { The black hexagon corresponds to Tully et al. 2013 data (see text). The open triangle, square and hexagon,  correspond to the estimates of  $\sigma_{**}$ (see text) from Ma \& Pan (2014), Ma et al. (2011) and Macaulay et al. (2012), respectively.} The filled triangle is the $\sigma_{\rm pec}$  corresponding to the Local Group bulk flow measurement (Kogut et al.\ 1993) assuming a Maxwell--Boltzmann distribution. The filled square corresponds to a $\Lambda$CDM nonlinear reconstruction of the Local Universe bulk flow (Hess \& Kitaura 2016). The discontinuous line is obtained from the linear $\Lambda$CDM WMAP9 predictions of the bulk flow (Carrick et al.\ 2015) and the shaded region shows the corresponding cosmic scatter in  $\sigma_{pec}$. The open circle is an individual velocity estimate (Mediavilla et al.\ 2015) from the lensed system Q2237+0305 (we have {located this point in $R$ according to the redshift of the lens galaxy}).}
\end{figure}

\begin{figure}[h]
\plotone{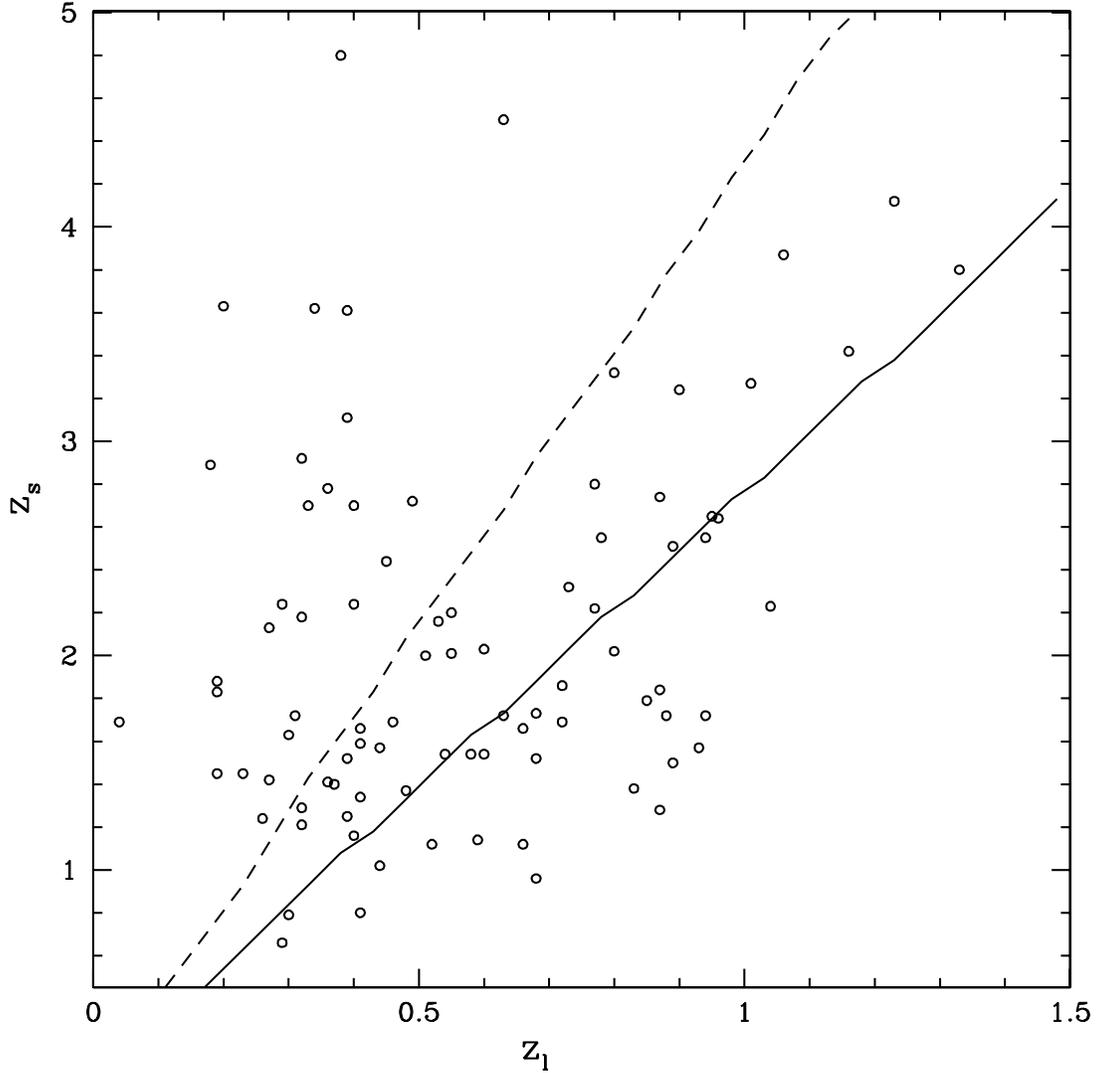}
\vskip -4 truecm
\caption{Circles represent the 2D distribution in redshifts ($z_l$,$z_s$) of the sample of gravitational lenses by Mosquera \& Kochanek (2011). Points above the continuous (dashed) curve have a $\hat \sigma_{\rm pec}(z_s)/\hat \sigma_{\rm pec}(z_l) $ estimate ratio below 0.1 (0.05). }
\end{figure}

\clearpage
\centering
\begin{deluxetable}{rccrrrcccc}
\rotate
\tablewidth{0pt}
\tablecaption{Light Curves Peaks}
\tablehead{
\colhead{Object  } &    image(s)     & \colhead{band}      &
\colhead{$\Delta m_- (\rm{S/N})$}          & \colhead{$\Delta m_+ (\rm{S/N})$}  &
\colhead{JD$-${ 2450000}}          & \colhead{$t_{eff}\,\rm (yr)$}    & \colhead{
}    &
 \colhead{comments}  &
\colhead{ref}\\
\colhead{$(z_l,z_s)$}}
\startdata

Q~0142-0100  & A-B & R &  0.15 (3)& 0.15 (3) & $\sim$2500& 10.3$\times$2& &POT&  (1,2)\\
 (0.49,2.72)& A-B & R &  0.05 (1)& 0.05 (1) & $\sim$3600& &  & & \\
\\

Q~J0158-4325 & A-B & r & $>$0.25 (5)& 0.35 (7) &3000  & 7.4$\times$2 & & POT& (3)  \\
(0.32,1.29)& A-B & r & $>$0.30 (6)& $>$0.25 (5) &$\sim$3500  &  & &  POT& \\
& A-B & r & {$>$0.12 (2})& $>$0.00 (0) &$\sim$3850  &  & & & \\
& A-B & r & $>$0.05 (1)& $>$0.11 (2) &$\sim$4600  &  & &  & \\
 & A-B & r & $$0.12 ({2})& $$0.28 ({6}) &4050  & & &      & \\
\\

HE~0435-1223 &A-B& R&$>$0.23 (4) & $>$0.30 (5) & $\sim$4250 & 7.9$\times$2 & & POT  & (4) \\
(0.46,1.69)& C-B & R & $$0.09 (2)  & $$0.09 (2) & 4750  &7.9& &&   \\
 & C-B & R & 0.14 (2)  & 0.27 (4) &   5600 & & & &  \\
 & D-B & R & 0.3 (3) & 0.3 (3)& 5600 &7.9  & & POT &  \\
\\

 
 SBS~0909+0532& { A-B}& g, { r} &$>$0.19 (6)  &$>$0.25 (8) & 4500 &7.81$\times$2& &POT& (5) \\ 
(0.83,1.38)&  { A-B} &  {g, r }&$>$0.13 (4)  &$>$0.08 (3) & 5700 & & & &  \\ 
\\
 
 RX J0911+0551 &B-A123 & I & \nodata  & \nodata & \nodata &3.01$\times 2$& &   & (6) \\
(0.77,2.80) \\
\\
 
 {SDSS 0924+0219}& A& R &  \nodata &\nodata  & \nodata  & 10.41 & &  & (7)  \\ 
(0.39,1.52)& B& R &  $>$0.15 (1) & $>$
0.6 (3)& 3100  & 10.41 & &  &  \\ 
& C& R &  \nodata &1.00 (5)  & $\lesssim$3000 & 10.41 & & &  \\ 
 & D& R &  \nodata &\nodata  & \nodata  & 10.41 & &  &  \\ 
  \\
  
 FBQ~0951+2635  & B-A&  R& \nodata  & \nodata   & \nodata   & 0.74 & &  & (8) \\   
(0.24,1.24) \\
\\
 
{Q~0957+0561}& B-A & V, R & 0.22 (3)& 0.26 (4) & -4500& 17.26$\times$2 & & POT & (9,10,11) \\ 
(0.36,1.41)& B-A & V, R & 0.08 (4) & $>$0.06 (3)& 5000&& & &  \\ 
   & &  &  & & & & & &  \\

 J~1001+5027& B-A & R & \nodata  & \nodata & \nodata & 6.7$\times 2$  & &   & (12)\\ 
(0.42,1.84)\\
\\
 
 SDSS~1004+4112& A-B &  r&\nodata  &\nodata & \nodata & 6.2$\times$2 & & & (13)\\ 
 (0.68,1.74 )& C-B &  r&  0.35 (4)& 0.4 (4)&3850 & 4.1 & &POT & \\ 
 & D-B &  r&  \nodata& \nodata& \nodata & 2.9 & & & \\ 
 \\
 
 SDSS J~1029+2623& B+C-A&  r & \nodata & \nodata & \nodata  &  3.42$\times 3$ & & &(14)\\ 
(0.55,2.20)\\
\\
 
{HE~1104-1805}& A-B& R & 0.25 (2)  & 0.25 (2) & 1800 & 13.7$\times 2$ & & & (15,16)\\ 
(0.73,2.32)\\
\\
 
 {RX~J1131-1231}& A-B& R &$>$
 0.4 (8) &$>$
 0.8 (14)  &$\sim$5100  & 8.46$\times 2$& &POT & (17)  \\ 
 (0.29,0.66)&D-B & R & \nodata & \nodata & \nodata & 8.18 & &&\\ 
 &C-B & R &  \nodata& \nodata& \nodata& 8.46 &   & \\ 
 \\
 
SDSS~J1206+4332&B-A & R &\nodata  & \nodata&  \nodata& 6.3$\times 2$ &  & & (18)  \\ 
(0.75,1.79)\\
  \\
 
 
 SDSS~J1353+1138&B-A& R & \nodata  &\nodata  &  \nodata & 0.68$\times 2$&  &  &(19)\\ 
(0.30,1.63)\\
\\

        

 
 
 WFI~J2033-4723&  A,B,C & R  &  \nodata  &  \nodata &  \nodata & 3.29$\times 3$ & &&(20) \\ 
(0.66,1.66) \\
\\


HS~2209+1914&B-A & R & \nodata & \nodata& \nodata& 8.28$\times 2$ & & & (18)\\ 
(0.5,1.07)\\
\\

    \enddata
 
   \tablerefs{(1) Oscoz et al 2013; (2) Koptelova et al. 2012; (3) Morgan et al. 2012; (4) Blackburne et al. 2014; (5) Hainline et al 2013; (6) Hjorth et al. 2002;  (7) MacLeod et al. 2015; (8) Jakobsson 2005; (9) Oscoz et al. 2002;  (10) Hainline et al. 2012; (11) Pelt et al. 1998; (12) Rathna Kumar et al. 2013; (13) Fiann et al. (2016); (14) Fohlmeister et al. 2013; (15) Poindexter et al. 2007; (16) Blackburne et al. 2015;  (17) Chartas et al. 2012; (18) Eulaers et al. 2013;  (19) Paraficz et al. 2009;  (20) Vuisoz et al. 2008.}

    \end{deluxetable}
    
\clearpage
\tiny
\centering
\begin{deluxetable}{ccc}
\tablewidth{0pt}
\tablecaption{$\sigma_{pec}$ Relative Error Budget}
\tablehead{
\colhead{Source } &    present ensemble   & \colhead{LSST survey}}

\startdata
$R$ & $< $0.10 & $< $0.05\tablenotemark{1}\cr
profile & $< $0.07 & $< $0.07 \cr
$\alpha$ & $< $0.04 & $< $0.04\tablenotemark{1} \cr
$\langle m \rangle$ & $<$0.15 & $<$0.15\cr
($\kappa,\gamma$) & $< $0.03 &  $< $0.03\tablenotemark{1}\cr
$\sigma_*$ & $< $0.00 & $< $0.00\cr
$v_0$ &  $< $0.00 & $< $0.00 \cr
$1/\sqrt{n}$ & 0.29 & 0.02$-$0.07 \cr
\enddata
\tablenotetext{1}{Overestimated upper limit}
\end{deluxetable}
    

\begin{thebibliography}{}

\bibitem[Assef et al.(2011)]{2011ApJ...742...93A} Assef, R.~J., Denney, K.~D., Kochanek, C.~S., et al.\ 2011, \apj, 742, 93 
\bibitem[B(2009)]{B} Blackburne, J.A. \ 2009, Ph.D. Thesis, Massachusetts Institute of Technology
\bibitem[Blackburne et al.(2014)]{2014ApJ...789..125B} Blackburne, J.~A., Kochanek, C.~S., Chen, B., Dai, X., \& Chartas, G.\ 2014, \apj, 789, 125 
\bibitem[Blackburne et al.(2015)]{2015ApJ...798...95B} Blackburne, J.~A., Kochanek, C.~S., Chen, B., Dai, X., \& Chartas, G.\ 2015, \apj, 798, 95 
\bibitem[Carrick et al.(2015)]{2015MNRAS.450..317C} Carrick, J., Turnbull, S.~J., Lavaux, G., \& Hudson, M.~J.\ 2015, \mnras, 450, 317 
\bibitem[Chang \& Refsdal(1979)]{1979Natur.282..561C} Chang, K., \& Refsdal, S.\ 1979, \nat, 282, 561 
\bibitem[Chang \& Refsdal(1984)]{1984A&A...132..168C} Chang, K., \& Refsdal, S.\ 1984, \aap, 132, 168
\bibitem[Chartas et al.(2012)]{2012ApJ...757..137C} Chartas, G., Kochanek, C.~S., Dai, X., et al.\ 2012, \apj, 757, 137 
\bibitem[de la Torre et al.(2013)]{2013A&A...557A..54D} de la Torre, S., Guzzo, L., Peacock, J.~A., et al.\ 2013, \aap, 557, A54 
\bibitem[Eulaers et al.(2013)]{2013A&A...553A.121E} Eulaers, E., Tewes, M., Magain, P., et al.\ 2013, \aap, 553, A121 
\bibitem[F(2009)]{F} Fiann, C. \ 2016, ApJ, submitted
\bibitem[Fohlmeister et al.(2013)]{2013ApJ...764..186F} Fohlmeister, J., Kochanek, C.~S., Falco, E.~E., et al.\ 2013, \apj, 764, 186 
\bibitem[Gil-Merino et al.(2005)]{2005A&A...432...83G} Gil-Merino, R., Wambsganss, J., Goicoechea, L.~J., \& Lewis, G.~F.\ 2005, \aap, 432, 83 
\bibitem[Hainline et al.(2012)]{2012ApJ...744..104H} Hainline, L.~J., Morgan, C.~W., Beach, J.~N., et al.\ 2012, \apj, 744, 104 
\bibitem[Hainline et al.(2013)]{2013ApJ...774...69H} Hainline, L.~J., Morgan, C.~W., MacLeod, C.~L., et al.\ 2013, \apj, 774, 69 
\bibitem[He{\ss} \& Kitaura(2016)]{2016MNRAS.456.4247H} He{\ss}, S., \& Kitaura, F.-S.\ 2016, \mnras, 456, 4247 
\bibitem[Hinshaw et al.(2009)]{2009ApJS..180..225H} Hinshaw, G., Weiland, J.~L., Hill, R.~S., et al.\ 2009, \apjs, 180, 225
\bibitem[Hjorth et al.(2002)]{2002ApJ...572L..11H} Hjorth, J., Burud, I., Jaunsen, A.~O., et al.\ 2002, \apjl, 572, L11 
\bibitem[Jakobsson et al.(2005)]{2005A&A...431..103J} Jakobsson, P., Hjorth, J., Burud, I., et al.\ 2005, \aap, 431, 103 
\bibitem[Jim{\'e}nez-Vicente et al.(2015a)]{2015ApJ...799..149J} Jim{\'e}nez-Vicente, J., Mediavilla, E., Kochanek, C.~S., \& Mu{\~n}oz, J.~A.\ 2015a, \apj, 799, 149
\bibitem[Jim{\'e}nez-Vicente et al.(2015b)]{2015ApJ...806..251J} Jim{\'e}nez-Vicente, J., Mediavilla, E., Kochanek, C.~S., \& Mu{\~n}oz, J.~A.\ 2015b, \apj, 806, 251 
\bibitem[Kaiser(1987)]{1987MNRAS.227....1K} Kaiser, N.\ 1987, \mnras, 227, 1 
\bibitem[Kaiser(1988)]{1988MNRAS.231..149K} Kaiser, N.\ 1988, \mnras, 231, 149 
\bibitem[Kayser et al.(1986)]{1986A&A...166...36K} Kayser, R., Refsdal, S., \& Stabell, R.\ 1986, \aap, 166, 36 
\bibitem[Kochanek(2004)]{2004ApJ...605...58K} Kochanek, C.~S.\ 2004, \apj, 605, 58
\bibitem[Kochanek et al.(1996)]{1996ApJ...473..610K} Kochanek, C.~S., Kolatt, T.~S., \& Bartelmann, M.\ 1996, \apj, 473, 610 
\bibitem[Koda et al.(2014)]{2014MNRAS.445.4267K} Koda, J., Blake, C., Davis, T., et al.\ 2014, \mnras, 445, 4267 
\bibitem[Kogut et al.(1993)]{1993ApJ...419....1K} Kogut, A., Lineweaver, C., Smoot, G.~F., et al.\ 1993, \apj, 419, 1 
\bibitem[Koptelova et al.(2012)]{2012A&A...544A..51K} Koptelova, E., Chen, W.~P., Chiueh, T., et al.\ 2012, \aap, 544, A51 
\bibitem[Kundic \& Wambsganss(1993)]{1993ApJ...404..455K} Kundic, T., \& Wambsganss, J.\ 1993, \apj, 404, 455 
\bibitem[Lavaux et al.(2013)]{2013MNRAS.430.1617L} Lavaux, G., Afshordi, N., \& Hudson, M.~J.\ 2013, \mnras, 430, 1617 
\bibitem[Li et al.(2006)]{2006MNRAS.368...37L} Li, C., Jing, Y.~P., Kauffmann, G., et al.\ 2006, \mnras, 368, 37 
\bibitem[Ma et al.(2011)]{2011PhRvD..83j3002M} Ma, Y.-Z., Gordon, C., \& Feldman, H.~A.\ 2011, \prd, 83, 103002 
\bibitem[Ma \& Pan(2014)]{2014MNRAS.437.1996M} Ma, Y.-Z., \& Pan, J.\ 2014, \mnras, 437, 1996 
\bibitem[Macaulay et al.(2012)]{2012MNRAS.425.1709M} Macaulay, E., Feldman, H.~A., Ferreira, P.~G., et al.\ 2012, \mnras, 425, 1709 
\bibitem[MacLeod et al.(2015)]{2015ApJ...806..258M} MacLeod, C.~L., Morgan, C.~W., Mosquera, A., et al.\ 2015, \apj, 806, 258 
\bibitem[Marshall et al.(2010)]{2010AAS...21540115M} Marshall, P.~J., Bradac, M., Chartas, G., et al.\ 2010, Bulletin of the American Astronomical Society, 42, 401.15 
\bibitem[Mediavilla et al.(2006)]{2006ApJ...653..942M} Mediavilla, E., Mu{\~n}oz, J.~A., Lopez, P., et al.\ 2006, \apj, 653, 942 
\bibitem[Mediavilla et al.(2011)]{2011ApJ...741...42M} Mediavilla, E., 
Mediavilla, T., Mu{\~n}oz, J.~A., et al.\ 2011, \apj, 741, 42 
\bibitem[Mediavilla et al.(2015)]{2015ApJ...798..138M} Mediavilla, E., Jimenez-Vicente, J., Mu{\~n}oz, J.~A., Mediavilla, T., \& Ariza, O.\ 2015, \apj, 798, 138 
\bibitem[Morgan et al.(2010)]{2010ApJ...712.1129M} Morgan, C.~W., Kochanek, C.~S., Morgan, N.~D., \& Falco, E.~E.\ 2010, \apj, 712, 1129 
\bibitem[Morgan et al.(2012)]{2012ApJ...756...52M} Morgan, C.~W., Hainline, L.~J., Chen, B., et al.\ 2012, \apj, 756, 52 
\bibitem[Mosquera \& Kochanek(2011)]{2011ApJ...738...96M} Mosquera, A.~M., \& Kochanek, C.~S.\ 2011, \apj, 738, 96 
\bibitem[Mosquera et al.(2013)]{2013ApJ...769...53M} Mosquera, A.~M., Kochanek, C.~S., Chen, B., et al.\ 2013, \apj, 769, 53 
\bibitem[Oguri et al.(2014)]{2014MNRAS.439.2494O} Oguri, M., Rusu, C.~E., \& Falco, E.~E.\ 2014, \mnras, 439, 2494 
\bibitem[Oscoz et al.(2002)]{2002ApJ...573L...1O} Oscoz, A., Alcalde, D., Serra-Ricart, M., Mediavilla, E., \& Mu{\~n}oz, J.~A.\ 2002, \apjl, 573, L1 
\bibitem[Oscoz et al.(2013)]{2013ApJ...779..144O} Oscoz, A., Serra-Ricart, M., Mediavilla, E., \& Mu{\~n}oz, J.~A.\ 2013, \apj, 779, 144 \bibitem[Paraficz et al.(2009)]{2009A&A...499..395P} Paraficz, D., Hjorth, J., \& El{\'{\i}}asd{\'o}ttir, {\'A}.\ 2009, \aap, 499, 395 
\bibitem[Peebles(1980)]{1980lssu.book.....P} Peebles, P.~J.~E.\ 1980, The Large Scale Structure of the Universe, Princeton University Press, Princeton
\bibitem[Pelt et al.(1998)]{1998A&A...336..829P} Pelt, J., Schild, R., Refsdal, S., \& Stabell, R.\ 1998, \aap, 336, 829 
\bibitem[Peng et al.(2006)]{2006ApJ...649..616P} Peng, C.~Y., Impey, C.~D., Rix, H.-W., et al.\ 2006, \apj, 649, 616 
\bibitem[Poindexter \& Kochanek(2010)]{2010ApJ...712..658P} Poindexter, S., \& Kochanek, C.~S.\ 2010, \apj, 712, 658 
\bibitem[Poindexter et al.(2007)]{2007ApJ...660..146P} Poindexter, S., Morgan, N., Kochanek, C.~S., \& Falco, E.~E.\ 2007, \apj, 660, 146 
\bibitem[Rathna Kumar et al.(2013)]{2013A&A...557A..44R} Rathna Kumar, S., Tewes, M., Stalin, C.~S., et al.\ 2013, \aap, 557, A44 
\bibitem[Scrimgeour et al.(2016)]{2016MNRAS.455..386S} Scrimgeour, M.~I., Davis, T.~M., Blake, C., et al.\ 2016, \mnras, 455, 386 
\bibitem[Shakura \& Sunyaev(1973)]{1973A&A....24..337S} Shakura, N.~I., \& Sunyaev, R.~A.\ 1973, \aap, 24, 337
\bibitem[Schechter et al.(2014)]{2014ApJ...793...96S} Schechter, P.~L., Pooley, D., Blackburne, J.~A., \& Wambsganss, J.\ 2014, \apj, 793, 96 
\bibitem[Sheth(1996)]{1996MNRAS.279.1310S} Sheth, R.~K.\ 1996, \mnras, 279, 1310 
\bibitem[Tinker et al.(2007)]{2007ApJ...659..877T} Tinker, J.~L., Norberg, P., Weinberg, D.~H., \& Warren, M.~S.\ 2007, \apj, 659, 877 
\bibitem[Treu et al.(2009)]{2009ApJ...690..670T} Treu, T., Gavazzi, R., Gorecki, A., et al.\ 2009, \apj, 690, 670 
\bibitem[Tully et al.(2013)]{2013AJ....146...86T} Tully, R.~B., Courtois, H.~M., Dolphin, A.~E., et al.\ 2013, \aj, 146, 86 
\bibitem[Vuissoz et al.(2008)]{2008A&A...488..481V} Vuissoz, C., Courbin, F., Sluse, D., et al.\ 2008, \aap, 488, 481 
\bibitem[Wambsganss(2006)]{2006glsw.conf..453W} Wambsganss, J.\ 2006, Saas-Fee Advanced Course 33: Gravitational Lensing: Strong, Weak and Micro, 453
\bibitem[Wyithe et al.(1999)]{1999MNRAS.309..261W} Wyithe, J.~S.~B., Webster, R.~L., \& Turner, E.~L.\ 1999, \mnras, 309, 261 
\bibitem[Wyithe et al.(2000a)]{2000MNRAS.315..337W} Wyithe, J.~S.~B., Webster, R.~L., \& Turner, E.~L.\ 2000a, \mnras, 315, 337 
\bibitem[Wyithe et al.(2000b)]{2000MNRAS.312..843W} Wyithe, J.~S.~B., Webster, R.~L., \& Turner, E.~L.\ 2000b, \mnras, 312, 843 

\end{thebibliography}
\end{document}